\documentclass[11pt]{amsart}  
\usepackage{amsmath,amsfonts,amssymb}
\usepackage{graphicx}

\newcommand{\be}[1]{\begin{equation}\label{#1}}
\newcommand{\ee}{\end{equation}}
\newcommand{\bea}[1]{\begin{eqnarray}\label{#1}}
\newcommand{\eea}{\end{eqnarray}}
\newcommand{\no}{\nonumber \\}

\newcommand{\Eq}[1]{Eq.(\ref{#1})}
\newcommand{\om}{\omega}
\newcommand{\Om}{\Omega}

\def\a0{{\alpha_0}}

\def\da0{{\dot{\alpha}_0}}

\newcommand{\half}{\frac{1}{2}}

\def\myoverDefn#1#2{\hbox{\space \raise-2mm\hbox{$\textstyle{#1} \atop \scriptstyle{#2}$} }}

\def\np{{n_p}}
\def\np0{{n_{p0}}}
\def\ns{n_{s}}
\def\ns0{n_{s0}}

\def\rvec{\mathbf{r}}
\def\nrvec{n(\mathbf{r})}
\def\Omrvec{\Omega(\mathbf{r})}
\def\Phirvec{\Phi(\mathbf{r})}
\def\nablabf{\boldsymbol{\nabla}}

\title[High-D Quantum Comm.]{Higher dimensional quantum communication in a curved spacetime:\\ an efficient simulation of the propagation of the wavefront of a photon}

\author{Warner A. Miller${}^1$, Paul M. Alsing${}^2$, and Doyeol Ahn${}^{1,3}$}
\address{${}^1$Department of Physics, Florida Atlantic University, Boca Raton, FL, 33431}
\address{${}^2$Air Force Research Laboratory, Information Directorate,Rome, NY 13441}
\address{${}^3$Center Quantum Information Processing, Department of Electrical and Computer Engineering,University of Seoul, Seoul 130-743, Republic of Korea}

\begin{document}

\begin{abstract}
A photon with a modulated wavefront can produce a quantum communication channel in a larger Hilbert space.  For example, higher dimensional quantum key distribution (HD-QKD) can encode information in the transverse linear momentum (LM) or orbital angular momentum (OAM) modes of a photon.  This is markedly different than using the intrinsic polarization of a photon.  HD-QKD has advantages for free space QKD since it can increase the communication channelÕs tolerance to bit error rate (BER) while maintaining or increasing the channels bandwidth. We describe an efficient numerical simulation of the propagation photon with an arbitrary complex wavefront in a material with an isotropic but inhomogeneous index of refraction.   We simulate the waveform propagation of an optical vortex in a volume holographic element in the paraxial approximation using an operator splitting method. We use this code to analyze an OAM volume-holographic sorter.  Furthermore, there are analogue models of the evolution of a wavefront in the curved spacetime environs of the Earth that can be constructed using an optical medium with a given index of refraction.  This can lead to a work-bench realization of a satellite HD-QKD system.
\end{abstract}

\maketitle


\section{INTRODUCTION}
\label{sec:intro}  

\begin{figure} [ht]
\centering
   \includegraphics[height=3.33 in]{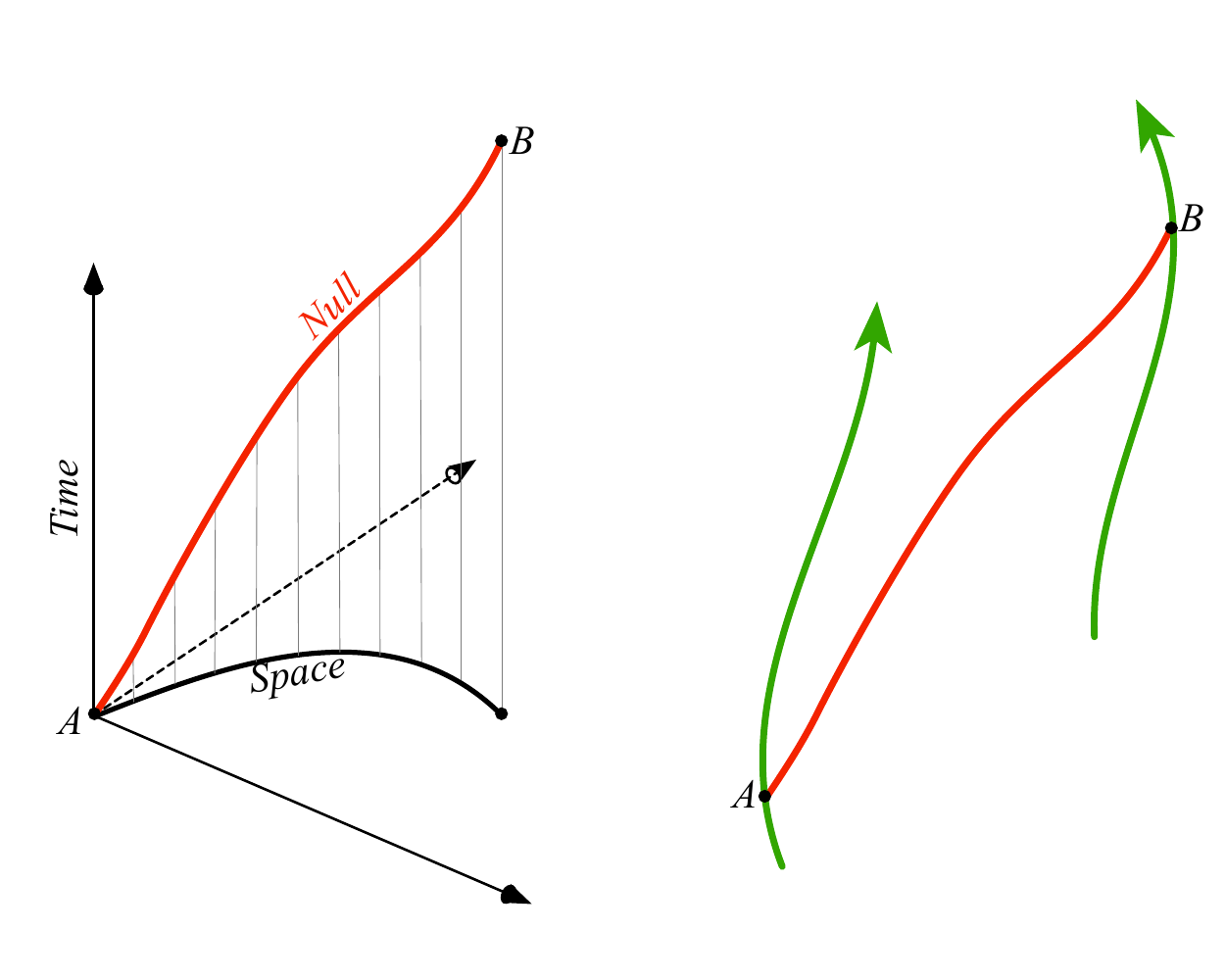}
   \caption{   
   \label{fig:ABST}
We show  Alice and Bob's time-like world lines connected by a photon at events $A$ and $B$; respectively.  The left diagram shows this null photon trajectory projected into the laboratory frame of Alice.  In her laboratory, the photon appears to bend under the influence of gravity.  The critical issue here is how to correctly describe the covariant propagation of the photon and its wavefront in Alice's 3-dimensional non-covariant laboratory. In Sec.~\ref{sec:n} we provide an optical-mechanical analogue of this propagation for conformally-flat spacetimes that describe this 3-dimensional wavefront propagation. Our paraxial approximations will be valid for small deviations of the photon in the spacelike hypersurface (black line labled space).}
   \end{figure}

We are interested in numerical tools to efficiently and accurately simulate a photon wavefront propagating in a mildly curved ($M/r\ll1$ ) spacetime geometry for the secure distribution of a one time only key from a sender Alice ($A$) to a receiver Bob ($B$), as illustrated in Fig.~\ref{fig:ABST}.  To secure such a key requires a quantum key distribution (QKD) system, and involves state preparation, state propagation, and state detection.  This QKD scenario has been exhaustively studied in the literature and is replete with security proofs for numerous protocols,
e.g.  \cite{Bennett:84,Ekert:91,Ekert:00}. These security proofs have been extended in many cases to higher-dimensional state spaces \cite{Cerf:02}, and all of the protocols have been or are currently being demonstrated successfully on earth, and in free-space links to satellites.\cite{Groblacher:06,Erhard2018}.

Conventional realizations of QKD today involve transmitting
heavily-attenuated laser pulses from Alice to Bob and encoding qubit
information in each packet by utilizing the spin of the photon. This
allows Alice and Bob, who are suitably authenticated, the possibility
to establish and share an arbitrarily-secure one-time only key between
them. Here they have access to a two-dimensional state space and can
therefore form three MUBs each with two orthogonal polarization
states. Such a six-state QKD scheme \cite{Bruss:98} has limited
bandwidth and optical fidelity constraints. These constraints can be
ameliorated by extending the QKD to higher-dimensional state space
\cite{Cerf:02}.

The potential of extending photon-based QKD to higher dimensions was first introduced in a delayed choice experiment using a photons orbital angular (OAM) in 1983 by Wheeler and Miller,  and was popularized in 1992 when Allen et~al. showed that
Laguerre-Gaussian light beams possessed a quantized orbital angular momentum (OAM) of $l\hbar$ per photon  \cite{Miller:84,Allen:92}. This opened up an arbitrarily high dimensional quantum space to a single photon \cite{Allen:03}. This was further developed by  Mair et~al. when they showed that  pairs of OAM photons can be entangled using parametric down
conversion, and then when  Molina-Terriza et~al. introduced a scheme to prepare photons
in multidimensional vector states of OAM  \cite{Mair:01,Oemrawsingh:04,Molina-Terriza:02} .   These early discoveries began the approach of OAM QKD.

There are advantages and disadvantages of using the extrinsic wavefront modulation of a photon to communicate quantum information as opposed to the intrinsic polarization states of a photon.  The main advantage is that higher-dimensional QKD can, in principle, 
increase bandwidth and be more tolerant to  bit error
rate (BER) \cite{Cerf:02}. It also offers a higher degree of entanglement. However, there are disadvantages as well.  First, such modulated photon wavefronts can be degraded in propagation due to variations in index of refraction, where a photon's polarization is more robust \cite{Paterson:05,Aksenov:08}  Second,  an apertured beam diverges. For OAM photons, this divergence is proportional to the square root of the angular momentum, and substantially larger apertures will be needed. Finally, one needs to efficiently sort OAM photons.  Given these difficulties, it has been pointed out that photon wavefronts that are a superposition of a certain set of transverse linear momentum states can be used as a signal for higher-dimensional QKD, and these states can be feasibly and efficiently sorted using volume holographic elements \cite{Miller:2011}.   

We focus here on the efficient and accurate simulation of the detection and propagation of single photons with
arbitrary complex wavefronts in a curved spacetime geometry.  In particular we simulate OAM modes or optical vortices. In Sec.~\ref{sec:pwe} we review the paraxial wave equation and its similarity to the Schr\"odinger equation. In this section, we also discuss the split operator method to solve the paraxial wave equation. This is the numerical approach we will use in our simulations. It is efficient and accurate as it utilizes the fast Fourier transform (FFT). In Sec.~\ref{sec:ovs} we simulate an optical vortex sorter using our split operator code. It involves the propagation of a wavefront through a isotropic but inhomogeneous optical medium. In Sec.~\ref{sec:n} we review an optical-mechanical analogue of a conformally flat spacetime. In this way we can simulate the propagation of our wavefronts in a curved spacetime through an equivalent optical medium with varying index of refraction. In Sec.~\ref{sec:pwe} we solve the propagation of a modulated wavefront in a gravitational field using the optical-mechanical analogue. The idea of simulating a gravitational field by an effective index of refraction has been around for a long time,\cite{Eddington:1920} and have recently come to the fore in quantum optics and quantum computing technologies.We solve the paraxial wave equation for this state for an equivalent index of refraction, and in Sec.~\ref{sec:fini} we discuss the results. 

\section{THE PARAXIAL WAVE EQUATION IN FLAT SPACETIME AND THE OPERATOR SPLITTING METHOD }
\label{sec:pwe}

The vector equation for the electric field in a flat spacetime in some inertial observers frame is obtained by the covariant Maxwell's equations where, 
\begin{equation}
\label{eq:vwe}
-\frac{n^2}{c^2} E_{,tt} = -\nabla^2 E - 2 \nabla \left(  E \cdot \nabla\left(\log{(n)}\right) \right).
\end{equation}
Here, we have assumed that the source-free medium is linear and isotropic, and that it is non magnetic, $\mu=\mu_0=const.$ However the medium may be inhomogeneous $\epsilon=\epsilon(r)$, as well as the index of refraction, 
\begin{equation}
n = \frac{\sqrt{\mu_0\epsilon}}{\sqrt{\mu_0\epsilon_0}}=\frac{c}{v},
\end{equation}
where we use geometric units where the speed of light,  $c=1$. Separating variables in Eq.~\ref{eq:vwe} with 
\begin{equation}
E(r,t)=u(r)T(t), 
\end{equation}
we obtain the usual modified Helmholtz equation for the spatial electric field,
\begin{equation}
\label{eq:mhe}
\nabla^2 u + \underbrace{2 \nabla \left(  u \cdot \nabla\left(\log{(n)}) \right)\right) }_{\approx 0} +n^2 k_0^2 u = 0,
\end{equation}
where $k_0=\omega_0/c$ is the vacuum wave number and $\omega_0$ is the vacuum angular frequency constant from the separation of variables, 
\begin{equation} 
T(t) \propto e^{i\omega_0 t}.
\end{equation} 
The under-braced term in Eq.~\ref{eq:mhe} is ordinarily non-zero for an inhomogeneous medium.  However, we can neglect this term for our purposes since our apertures are large with respect to our wavelength, and since we are dealing with slowly-varying and small index of refraction variations in a thick dielectric diffraction grating (volume hologram).  If we neglect this term then we obtain the usual Helmholtz, or scalar wave equation, 
\begin{equation}
\label{eq:he}
\nabla^2 u + k^2 u = 0,
\end{equation}
where the wave number $k=n k_0$ depends on the inhomogeneous isotropic index of refraction of the medium we are propagating through. 
This equation can be solved efficiently and accurately if we make one more assumption that applied to our analysis in this manuscript. In particular, we make the slowly-varying envelope approximation, where we assume the electric field is propagating along the $z$ axis in our observers laboratory frame and the remaining field amplitude is a slowly varying function, $\psi(x,y,z)$.  Here we rewritie Eq.~\ref{eq:he} 
\begin{equation}
\label{eq:hez}
\left(\frac{\partial^2}{\partial z^2} + \nabla_T^2 \right) u + k^2 u =0,
\end{equation}
where $\nabla_T$ is the transverse Laplacian,
\[
\nabla_T^2 := \left(\frac{\partial^2}{\partial x^2}\right) +\left(\frac{\partial^2}{\partial y^2}\right)
\]
and 
\begin{equation} 
\label{eq:psi}
u = \psi\, e^{i k_0 z}.
\end{equation}
Inserting, Eq.~\ref{eq:psi} into Eq.~\ref{eq:hez} yields
\begin{equation}
\left(   
\underbrace{\frac{\partial^2 \psi}{\partial z^2}}_{\approx 0} +2 i k_0 \frac{\partial \psi}{\partial z} -k_0^2 \psi + \nabla_T^2 \psi 
\right) 
+ k^2 \psi = 0.
\end{equation}
At this point, we assume that the wave vector is close to the optical axis, $z$, so 
\begin{equation}
\label{eq:paa}
\left| \frac{\partial^2 \psi}{\partial z^2} \right| \ll \left| k_0 \frac{\partial \psi}{\partial z}   \right|.
\end{equation}
This is referred to as the paraxial approximation in wave optics \cite{Saleh:2007}, resulting in a parabolic equation, 
\begin{equation}
2 i k_0 \frac{\partial \psi}{\partial z} = - \nabla_T^2 \psi + \left( k_0^2 -k^2\right) \psi = 0.
\end{equation}
This paraxial equation for optics is a 2-dimensional Schr\"odingers equation for a quantum particle in a potential $V$,
\begin{equation}
\label{eq:sepe}
\boxed{
\begin{array}{lrl}
\hbox{Paraxial Optics:}& -\frac{1}{i} \frac{\partial \psi}{\partial z}  &= -\frac{1}{2k_0} \nabla_T^2 \psi + V \psi,\\
&\\
\hbox{Quantum Mechanics:}&-\frac{\hbar}{i} \frac{\partial \psi}{\partial t}  &= -\frac{\hbar^2}{2m} \nabla^2 \psi + V \psi,\\
\end{array}
}
\end{equation}
where the effective optical potential depends on the index of refraction of the medium, 
\begin{equation}
\label{eq:ep}
V = \frac{1}{2} \left( k_0^2 -k^2\right) =  \frac{1}{2} k_0^2 \left( 1 -n^2\right), 
\end{equation}
and we make the identifications,
\begin{equation}
\label{eq:transforms}
\begin{array}{rll}
\hbox{QM} & {\ }& \hbox{PO}\\
\hline
\hbar & \longrightarrow & 1\\
m & \longrightarrow & k_0\\
t & \longrightarrow & z\\
\end{array}
\end{equation}

Fortunately, there are highly efficient numerical approaches to solve time-dependent Schr\"odinger-type equations.  We know of no better numerical algorithm for our purposes than the split operator method \cite{Feit:1980,Feit:1982,Hermann:1995}.  This method was developed initially to solve the paraxial optics equation, and later it was applied to the Schr\"odinger equation.  This particular approach uses the fast Fourier transform (FFT) , is second or higher order (or higher) convergent, and preserves normalization throughout the evolution. We briefly outline this algorithm.

We find it convenient to express the Hamiltonian operator in Eq.~\ref{eq:sepe} in terms of the sum of two operators, 
\begin{equation} 
\label{eq:teo}
-\frac{\hbar}{i} \frac{\partial}{\partial t} = H \psi = (T + V) \psi,
\end{equation}
where the momentum or ``slip operator,''
\begin{equation}
\label{eq:slip}
T :=  -\frac{\hbar^2}{2m} \nabla^2,
\end{equation}
and the ``kick operator," $V$.  We seek a unitary transformation operator that is a solution of Eq.~\ref{eq:teo},  therefore
\begin{equation}
\label{eq:ueo}
\psi(r,t) = {\mathcal U}\left(t,t_0\right) \psi(r,t_0) \Longrightarrow {\mathcal U}\left(t,t_0\right)= e^{-\frac{i}{\hbar} \int_{t_0}^{t} H dt}.
\end{equation}
This evolution operator can be subtile when there is time dependence (we will not consider this here), and care must be taken because of non-commutitivity of exponentiated operators. One must utilize the Campbell-Baker-Housdorff identity where for two operators $A$ and $B$, 
\begin{equation}
\label{eq:cbh} 
e^{A+B} = e^{A} e^{B} e^{-[A,B]/2}, 
\end{equation} 
is exact only if $[A,[A,B]]=  [B,[A,B]]=0$. One can use this identity to remove lower-order error terms in the unitary evolution operator to obtain a second-order accurate operator,
\begin{equation}
\label{eq:abba}
e^{H (t-t_0)} = e^{A/2+B+A/2} \approx e^{A/2} e^{B} e^{A/2}, 
\end{equation}. 
with 
\begin{align}
A &:= -i/\hbar T (t-t_0),\\
B &:= -i/\hbar V (t-t_0). \label{eq:2nd} 
\end{align}
If we consider a small time step where $t\rightarrow t_0+\delta t$ then the evolution from Eqs.\ref{eq:ueo}-\ref{eq:2nd}, then 
\begin{eqnarray}
\psi\left(r,t_0+\Delta t\right) & \approx & 
\underbrace{e^{- \frac{i}{\hbar}V\frac{\Delta t}{2}}}_{\frac{1}{2}\ kick}
\underbrace{e^{- \frac{i}{\hbar}T\Delta t}}_{slip}
\underbrace{e^{- \frac{i}{\hbar}V\frac{\Delta t}{2}}}_{\frac{1}{2}\ kick} \psi(r,t_0)\\
& \approx &  
\underbrace{e^{- \frac{i}{\hbar}T\frac{\Delta t}{2}}}_{\frac{1}{2}\ slip}
\underbrace{e^{- \frac{i}{\hbar}V\Delta t}}_{kick}
\underbrace{e^{- \frac{i}{\hbar}T\frac{\Delta t}{2}}}_{\frac{1}{2}\ slip}  \psi(r,t_0).
\end{eqnarray}
 These ``kick'' and ``slip'' operators can be determined numerically by a succession of  FFT, and inverse FFT in the following five steps:
 \begin{enumerate}
 \item Multiply to define, $\psi_1(r,t_0)  = e^{-i \frac{V}{2 \hbar}\Delta t}   \psi(r,t_0)$,
 \item Fourier transform to get $\phi_1(p,t_0) = {\mathcal F}\left(  \psi_1(r,t_0) \right)$,
 \item Multiply to generate  $\phi_2(p,t_0) = e^{-i \frac{T(p)}{\hbar}\Delta t} \phi_1(p,t_0)$,
 \item Inverse Fourier transform to obtain $\psi_2(r,t_0) = {\mathcal F}^{-1}\left(  \phi_1(p,t_0) \right)$, \ finally
 \item Multiply to define, $\psi(r,t_0+\Delta t)  = e^{-i \frac{V}{2 \hbar}\Delta t}   \psi_2(r,t_0)$.
 \end{enumerate}
One need not worry about the normalization factors that accompany the FFT since there is always an even number of FFT'a and inverse FFT's in the split operator algorithm. By making use of the transformations of Eq.~\ref{eq:transforms}, this split operator method applies equally well to the paraxial wave equation and its evolution, and this optical application is the primary focus of this manuscript. 

Before we consider optical wavefront propagation in a curved spacetime geometry, we will use the split operator method to analyze the propagation of an optical vortex through a volume hologram with inhomogeneous index of refraction in the next section. And in particular we will numerically calculate the efficiency of a volume-holographic optical-vortex sorter.  

\section{THIS VORTEX THAT VORTEX: AN OPTICAL VORTEX SORTER}
\label{sec:ovs}
\begin{figure} [ht]
\centering
   \includegraphics[height=2.5 in]{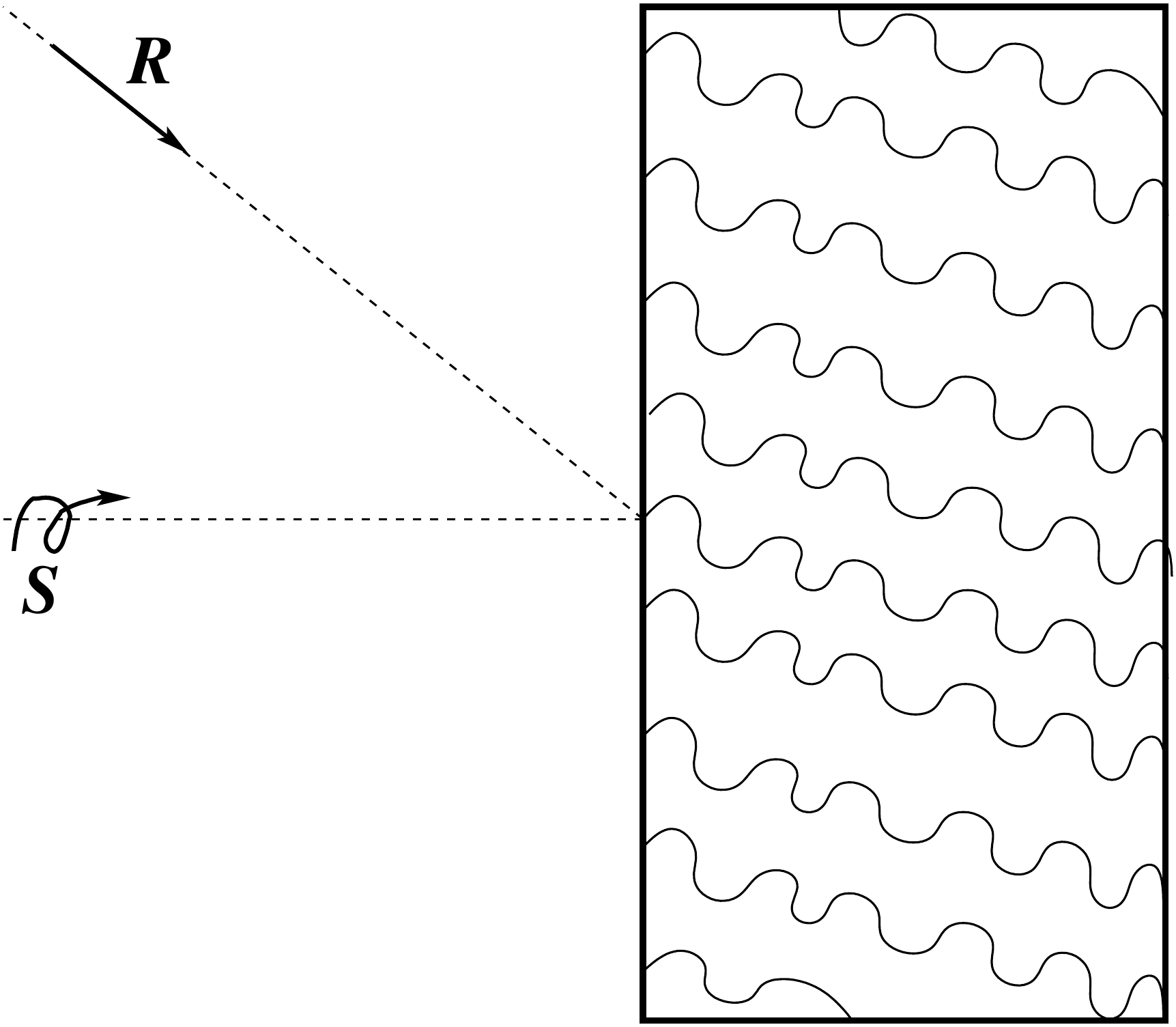}
   \caption
   { \label{fig:hologram}
The OAM volume Bragg grating produced by the interference of an optical vortex signal (S) and a reference plane wave (R). }
   \end{figure}

It is sufficient here to thoroughly analyze a single optical vortex grating and to exhibit its ability to differentiate a photon with one value of OAM from another. Once this has been demonstrated, the higher-dimensional sorters can be constructed by literally stacking or sequencing such gratings, one from each basis state in a given MUB. Another approach will be incoherently multiplex each of the gratings into a single holographic substrate. In this section,we examine a holographic emulsion produced by the coherent interference of a unit-amplitude plane wave and the unit amplitude OAM wave front corresponding to $\ell$.  Specifically, the hologram's face is oriented so that its normal is along the $z$-axis of our observer. The hologram's face or aperture is rectangular with dimensions $L_x$ by $L_y$ and the thickness of the hologram is $L_z$. The bulk index of refraction is $n_0$, but this is modulated by an interference pattern of amplitude $\Delta n$. The hologram is generated by the interference pattern of two waves of angular frequency $\omega_0$, (1) an incident reference plane wave whose wave vector, $k_r$, is in the $x$-$z$ plane and tilted at an angle $\theta_r$ from the $z$-axis, and (2) a modulated signal plane wave of wave vector, $k_s$,  also in the $x$-$z$ plane but titled an angle $\theta_s$ from the $z$-axis.  The modulation is  an optical vortex with orbital angular momentum $\ell$.  The interference of these two waves gives rise to a modulation in the index of refraction in the volume hologram,
\begin{equation}
n(x,y,x) = n_0 + \Delta n \cos{\left( (k_s -k_r)\cdot \rho +\ell\phi  \right)},
\end{equation} 
where, $\rho^2=x^2+y^2$ is the radius and $\phi$ is the azimuthal angle in the observer's cylindrical coordinates, $\{\rho,\phi,z\}$. This is illustrated in Fig.~\ref{fig:hologram}.
We can examine the sorting efficiency of this hologram by (1) sending in an identical signal wave with a Gaussian envelope  into the hologram and measuring how much of the signal is directed toward the detector in the far-field behind the hologram at an angle $\theta_r$ in the 
$x$-$z$-plane, and compare this with  (2) the signal intensity for a beam with a miss-matched value for its orbital angular momentum $\ell_s\ne \ell$.  

The initial amplitude of the optical wavefront in Eq.~\ref{eq:psi} i is a Gaussian beam of width $\sigma$ that is modulated by an optical vortex with orbital angular momentum, $\ell$, 
\begin{equation}
\label{eq:initialbeam}
u(x,y,z_0,t_0)= \psi(x,y,z_0,t_0) e^{ik_0z}= e^{-\rho^2/\sigma^2+ i \ell \phi}e^{ik_0z}.
\end{equation}

   \begin{figure} [ht]
\centering
   \includegraphics[height=2.5 in]{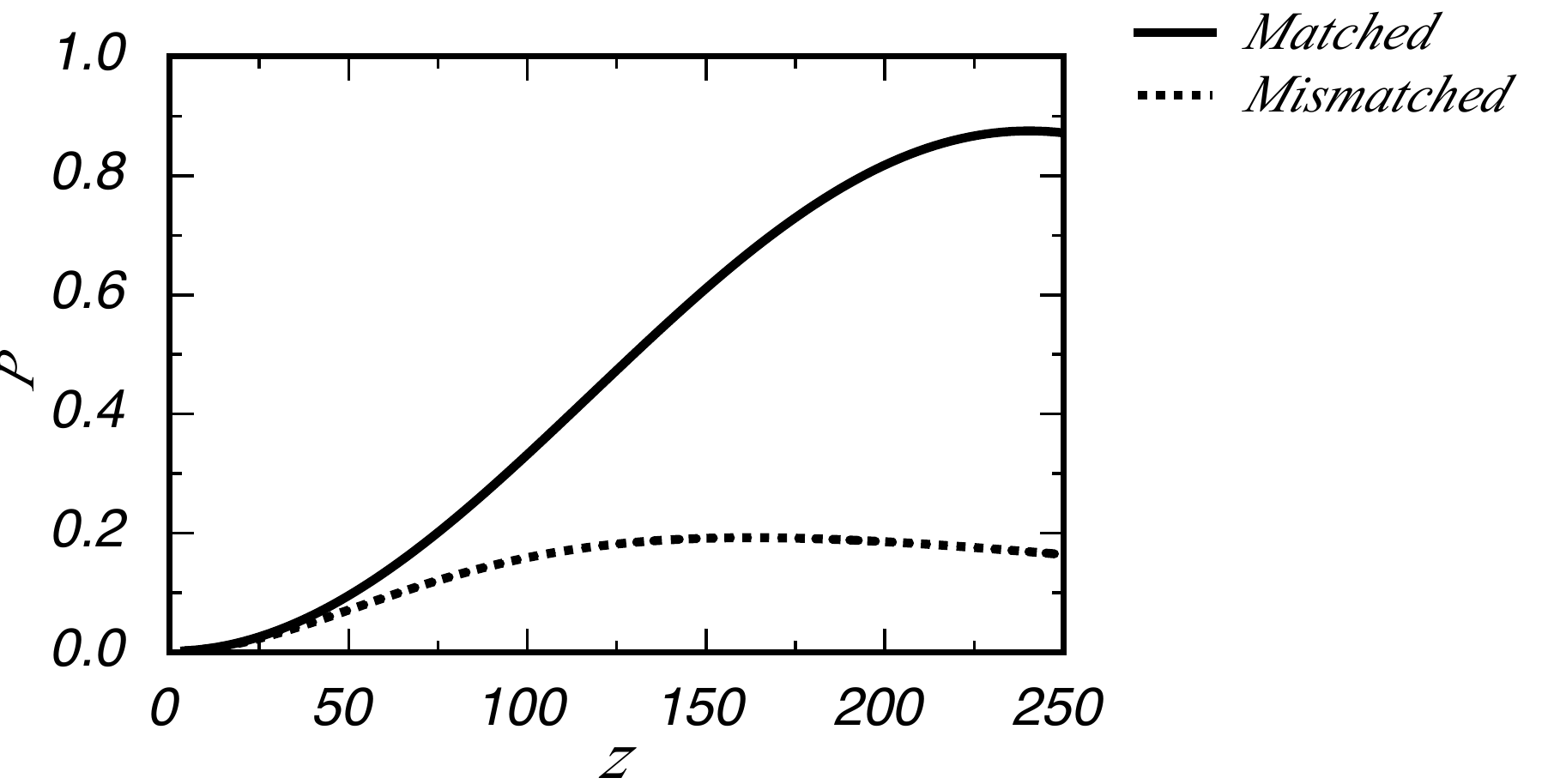}
   \caption
   { \label{fig:eff}
   We simulate the sorting efficiency of a hologram recorded with an $\ell=10$ optical vortex modulated by a Gaussian wave packet with $\sigma=200$. We plot the probability $P$ of a matched  $\ell=10$ signal diffracted into the detector (solid line) verses an $\ell=-10$ mismatched signal (dashed line).  The maximum sorting efficiency, $P\approx70.7\%$  is achieved at $z\approx245.44\lambda$.  In this simulation we set $k_0=2\pi$ so $\lambda=1$.  The angle between the reference beam and the signal beam in the recording of the hologram was $\theta=\pi/14$, so that the two wave vectors for the signal and reference were $k_s=\{0,0,1\}$ and $k_r=\{\sin(\theta) k_0,0,\cos(\theta) k_0\}$, respectively. We took the bulk index of refraction $n_0=1$, and the modulation to be $\Delta n=0.002$. The face of the hologram a square of length $400\lambda$, while the thickness was measured in $z$.  We simulated the efficiency for variable thicknesses $z\in[0,250]$.  We discretized the hologram by $512\times512\times512$ grid points.  }
   \end{figure}

\section{EFFECTIVE INDEX OF REFRACTION FOR A BLACK HOLE SPACETIME}
\label{sec:n}
Metrics in general relativity (GR) that can be written in isotropic coordinates lend themselves to a simple exact formulation such that the effects of the curved spacetime are encapsulated in a spatially varying
index of refraction $\nrvec$  \cite{Alsing:AJP:2001}. A metric in isotropic coordinates has the form
\begin{equation}
\label{eq:metric}
ds^2 = \Om^2(\rvec)\,c_0^2\,dt^2 - \Phi^{-2}(\rvec)\,d\rvec^2, 
\end{equation}
where $\rvec=(x,y,z)= \{x^i\}_{i\in\{1,2,3\}}$ are Cartesian coordinates. Thus, a metric in isotropic coordinates is conformally flat.
The isotropic coordinate speed of light $c(\rvec)$ is determined by the condition that the geodesics are null ($ds=0)$ leading to 
\begin{equation}
c(\rvec)=|d\rvec/dt| = c_0\,\Phirvec\,\Omrvec. 
\end{equation}
Thus the effective index of refraction for light in the gravitational field is then 
\begin{equation}
\nrvec = \Phirvec^{-1}\,\Omrvec^{-1}.
\end{equation}
The index of refraction may be used to formulate a Newtonian-like ``F=ma" formulation of geometric optics in which the equation for the path of the ray assumes the form of a Newton's law of motion \cite{Evans:1986,Evans:1996}
\begin{equation}
d^2\rvec/dA^2 = \nablabf\left(\half\,\nrvec^2\,c_0^2\right)
\end{equation}
 where $dA = dt/\nrvec^2$ is the so called optical action \cite{Alsing:AJP:2001}. The lefthand side of this equation has the form of an acceleration, while the righthand side has the form of a force, the gradient of a potential-energy function $U(\rvec) = -(\half\,\nrvec^2\,c_0^2)$. Using the optical action, one can also show that motion of massive particles is given by a similar equation,
\begin{equation}
d^2\rvec/dA^2 = \nablabf\left(\half\,n^2(\rvec)\,V^2(\rvec)\right)
\end{equation}
where 
\begin{equation}
V(\rvec) = |d\rvec/dt|= c_0/\nrvec. 
\end{equation}
Thus, in the massless limit, $V\rightarrow c_0$ this formula reduces to the one for light. Thus the latter formulas holds for \textit{both} massive and massless particles. While in this work we will be primarily interested in massless particles (light), in this section we will carry out a single formalism that is valid for both massive and massless particles.

The optical-mechanical analogy states that for light, the normal to surfaces of constant phase is the wavevector, while for massive particles, the normal to surfaces of constant action is the particle's momentum. This can put on a firm footing by considering a variational principle for the particle's geodesics via 
\begin{equation}
\delta \int_{\rvec_1,t_1}^{\rvec_2,t_2}\,ds =0
\end{equation}
using the line element $ds$ which contains the metric coefficients defining the spacetime. This can be cast into a Lagrangian form
\begin{equation}
\delta \int_{t_1}^{t_2}\,L(x^i,V^i)=0
\end{equation}
where
\begin{equation}
L(x^i, V^i) = -m\,c_0^2\,\Omega[1 - V^2\,n^2/c_0^2]^{1/2}.
\end{equation}
Here, $V^2 = \sum_{i=1}^{3} (V^i)^2$ with $V^i = d x^i/dt$
the components of the coordinate velocity. The canonical momenta are given by
\begin{equation}
p_i = \partial L/\partial x^i = -m\,\Omega\,n^2\,[1 - V^2\,n^2/c_0^2]^{-1/2}\,V_i,
\end{equation}
 and the effective Hamiltonian (energy) when expressed in terms of the momenta is given by 
 \begin{equation}
 H = m\,c_0^2\,[\Omega^2 + p^2/n^2\,m^2\,c_0^2]^{1/2}
 \end{equation}
which is constant of the motion along the geodesic. Lastly, the magnitude of the coordinate velocity is given by 
\begin{equation}
V(\rvec) = c_0\,n^{-1}(\rvec)\,[1 - m^2\,c_0^2\,\Omrvec/H]^{1/2}
\end{equation}
which generalizes the usual phase velocity for light $v=c_0/n$.

To bring in quantum mechanics, we follow \cite{Alsing:AJP:2001} de Broglie and define the constant coordinate frequency $\om$ via
$H\equiv \hbar\,\om$. The measurable proper frequency $\tilde{\om}$ along the particles geodesic is given by using the isotropic metric to obtain $\tilde{\om}(\rvec) = \om/\Omrvec$. By  using the de Broglie relationship $p(\rvec) \equiv \hbar\,k(\rvec)$ we obtain the coordinate wavelength $\lambda = \hbar\,c_0^2/(n^2(\rvec)\,H\,V(\rvec))$ with measurable proper
wavelength $\tilde{\lambda}(\rvec) = \lambda/\Phirvec$. One can then define an index of refraction $N(\rvec)$ for both massive and massless particles that satisfies $\lambda(\rvec)\,N(\rvec)=$ constant with
$N(\rvec)\equiv n\,[1-m^2\,c_0^4\,\Om^2(\rvec)/H^2]^{1/2}$,
which reduces to $N(\rvec)\rightarrow n(\rvec)$ for light by setting $m=0$.
We obtain gravitation redshifts by using $N(\rvec_1)\,\lambda(\rvec_1) = N(\rvec_2)\,\lambda(\rvec_2)$. Converting to proper wavelengths we have
$N(\rvec_1)\,\tilde{\lambda}(\rvec_1)\,\Phi(\rvec_1) = N(\rvec_2)\,\tilde{\lambda}(\rvec_2)\,\Phi(\rvec_2)$.
We can define the phase velocity $v_p = \om/k = H/p$, or equivalently $v_p(\rvec) = c_0/N(\rvec)$, which shows that $N(\rvec)$ is playing the proper role as an index of refraction.
By substituting the de Broglie definitions into the Hamiltonian we obtain 
$\hbar^2\,\om^2= m^2\,c_0^2\,\Om^2(\rvec) + c_0^2\,\hbar^2\,k^2(\rvec)/n^2(\rvec)$. Upon differentiating this expression with respect to $k$ we obtain the group velocity
$v_p(\rvec) = d\om/dk = c_0 N(\rvec)/n^2(\rvec)$, and with a little algebra this yields $v_g(\rvec) = V(\rvec)$, the magnitude of the coordinate velocity.

Finally, we can \textit{first quantize} the above Hamiltonian by converting it to a relativistic Klein-Gordon equation for the wavefunction $\psi(x)$ for a  spinless massive particle with the substitution of $p_\mu \rightarrow -i\,\hbar\,\nabla_\mu$ into the curved-spacetime mass-shell constraint 
\begin{equation}
g^{\mu\nu}\,p_\mu\,p_\nu - (m\,c_0)^2 = 0,
\end{equation}
 where $\nabla_\mu$ is the usual covariant derivative,
\begin{equation}
\nabla_\mu\,T_\nu(x) = \partial_\mu\,T_\mu(x) - \Gamma^\lambda_{\mu\,\nu}\,T_\lambda(x)
\end{equation}
with Christoffel connection
\begin{equation}
\Gamma^\lambda_{\mu\,\nu} = \half\,g^{\lambda\kappa}\,(\partial_\mu\,g_{\kappa\nu} + \partial_\nu\,g_{\mu\kappa}- \partial_\kappa\,g_{\mu\nu}).
\end{equation}
Thus, the Klein Gordon equation 
\begin{equation}
(g^{\mu\nu}\,\nabla_\mu\,\nabla_\nu + m^2\,c_0^2/\hbar^2)\,\psi(x)=0
\end{equation}
 becomes 
 \begin{equation}
 (g^{\mu\nu}\,\partial_\mu\,\partial_\nu -g^{\mu\nu}\,\Gamma^\lambda_{\mu\,\nu}\,\partial_\lambda + m^2\,c_0^2/\hbar^2)\,\psi(x)=0.
 \end{equation}
In a static isotropic metric $
g_{\mu\nu}(x) = \textrm{diag}(\Om^2\,c_0^2,-\Phi^{-2},-\Phi^{-2},-\Phi^{-2})$, and 
with the aid of the identity $g^{\mu\nu}\,\Gamma^\lambda_{\mu\,\nu} = -(-g)^{-1/2}\,\partial_\mu[(-g)^{1/2}\,g^{\mu\lambda}]$ we obtain 
\begin{equation}
\label{eq:yyy}
(n^2/c_0^2)\,\partial_t^2\psi = -\nablabf^2\,\psi - \nablabf\xi\cdot\nablabf\psi + k_c^2(\rvec)\,\psi=0,
\end{equation}
 where $\nablabf$ is the Cartesian divergence and $\nablabf^2$ is the Cartesian Laplacian. In Eq.~\ref{eq:yyy} we have defined $\xi(\rvec) = \ln(\Omrvec/\Phirvec)$ and $k_c^2(\rvec) = m\,c_0/(\hbar\,\Phirvec)$ which can be interpreted as the particle's coordinate Compton wavelength (the particle's invariant proper Compton wavelength is $h/(m\,c_0)$). We can remove the cross term $\nablabf\xi\cdot\nablabf\psi$ arising from the covariant derivative by letting $\psi(\rvec) \equiv f(\rvec)\,\phi(\rvec,t)$ where $f(\rvec) = (\Om/\Phi)^{1/2} \equiv e^{-\xi/2}$. Then $\phi(\rvec,t)$ satisfies the wave equation
$(n^2/c_0^2)\,\partial_t^2\phi - \nablabf^2\phi + [k_c^2(\rvec) + \eta(\rvec)] =0$ where
$\eta(\rvec) = \half\,\nabla^2\xi(\rvec)+ \frac{1}{4}\,|\nabla\xi(\rvec)|^2$. 
Finally, letting $\phi(\rvec,t) \equiv e^{i\,H\,t/\hbar}u(\rvec)$ we arrive at the Helmholz equation
valid for both massive and massless particles,
\be{Helmholz:massive:massless}
\nablabf^2\,u(\rvec) + [k^2(\rvec) - \eta(\rvec)]\,u(\rvec) = 0, 
\qquad k(\rvec)=k_0\,N(\rvec) = k_0\,\nrvec\,\sqrt{1-\dfrac{m^2 c_0^2\,\Omrvec^2}{\hbar^2\,\om_0^2}}, 
\ee
where we have used $H\equiv \hbar\,\om_0$ and $k_0=\om_0/c_0=H/(\hbar\,c_0)$.

For the Schwarzschild metric in isotropic coordinates we have
\bea{Sch:metric:iso}
\Om &=& \dfrac{1-1/\rho}{1+1/\rho}, \qquad\qquad \Phi = \dfrac{1}{(1+1/\rho)^2}, 
\qquad\qquad n =  \frac{1}{\Phi\,\Om} =  \dfrac{\;(1+1/\rho)^3}{(1-1/\rho)} \no
\xi &=& \ln(1-1/\rho^2),            \qquad  k = k_0\,n\,\sqrt{1-\Om^2/H^{'2}} 
\eea
where $H^{'}\equiv H/(m\,c_0^2)$ and the radial coordinate is given by $\rho=r/r_s$ with $r_s = G\,M/c_0^2$ (half the Schwarzschild radius $r_S = 2\,G\,M/c_0^2$). The constant $H^{'}$ can be evaluated at any point on the particle's geodesic where it's coordinate velocity is known. 

It can be shown \cite{Alsing:AJP:2001} that $\eta(\rvec)\ll k^2(\rvec)$ for all values of $1\le\rho\le\infty$ (the valid range of the isotropic scaled radius $\rho$), so that we can drop the former.
For massless particles $H'\rightarrow\infty$ ($m\rightarrow 0$) and $k=k_0\,n$, so that for light we end up with the Helmholz equation for a scalar field with a spatially varying wavevector $k(\rvec)$, 
\be{Helmholz:massless}
\nablabf^2\,u(\rvec) + k^2(\rvec) \,u(\rvec) = 0,
\qquad k(\rvec)=k_0\,n(\rvec) = \dfrac{k_0}{\Omrvec\,\Phirvec}, \qquad \textrm{(massless particles)}.
\ee
The above derivations reveal that the complete effects of the gravitational field are now encapsulated in the spatially varying index of refraction $\nrvec$ (for light) given by the combination of the metric coefficients $(\Omrvec\,\Phirvec)^{-1}$. Thus, for example, the well known bending of light around a massive object can now be seen as the usual bending of light as the photon moves into a more intense index of refraction \cite{Alsing:AJP:2001,Alsing:GRG:2001}. The fact the matter waves bend more than light is now attributed to the latter's slower velocity. The gravitational wave optics of massive and massless particles differ only in a dispersion effect, since the index of refraction  for massive objects $N(\rvec)$ depends on the parameter $m\,c_0^2/H$.

In the next section we will explore the paraxial approximation to \Eq{Helmholz:massless} for light and explore the propagation of wavefronts in a gravitational field described by the static Schwarzschild metric.

\section{PARAXIAL WAVE EQUATION IN A CURVED SPACETIME}\label{sec:pwegr}

\begin{figure} [ht]
\centering
   \includegraphics[height=2.5 in]{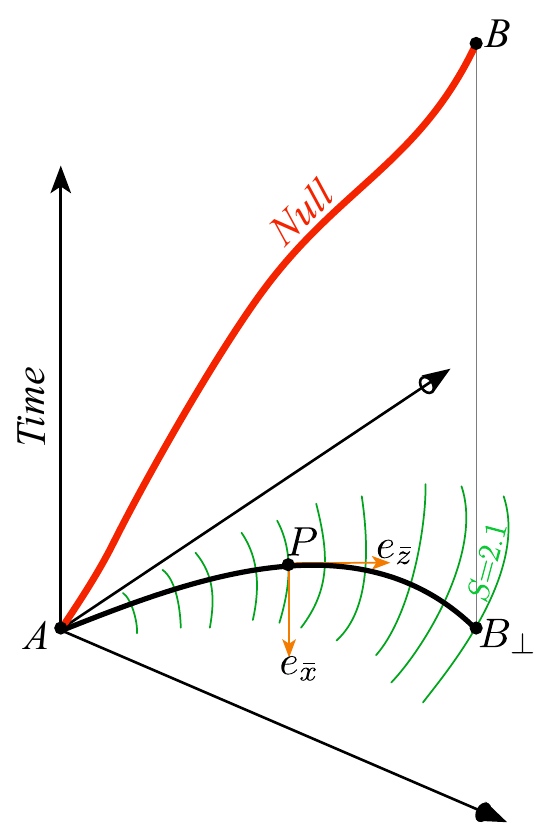}
   \caption
   { \label{fig:ABSTP}
We show  Alice and Bob's time-like world lines connected by a photon at events $A$ and $B$; respectively.  We suppress one spatial dimension for ease of visualization. This diagram shows this null photon trajectory projected into the laboratory frame of Alice.  In her laboratory, the photon appears to bend under the influence of gravity.}
   \end{figure}

In this section we will assume that both Alice and Bob are at rest in the isotropic coordinates $\{t,x,y,z\}$ of our conformally-flat spacetime of Eq.~\ref{eq:metric}. This assumption we make is not out of necessity, but will avoid complications of boosting into Alice and Bob's frame at event $A$ and $B$ shown in Fig.~\ref{fig:ABST}. The photon trajectory connecting $A$ to $B$ is determined by the wave 4-vector,
\begin{equation}
{}^{(4)}\!\nabla_k k = 0,
\end{equation}  
and in the optical-mechanical model we are following in Evans et al. the spatial projection of this null trajectory onto the $t=const$ spacelike hypersurface of Allice is given by their force equation,
\begin{equation}
\label{eq:force} 
\frac{d^2 r}{d A^2} = \nabla \left(\frac{1}{2} n^2 c_0^2\right),
\end{equation}
where $dA=dt/n^2$ and $\nabla={}^{\!(3)}\nabla$ is the 3-dimensional gradient operator. This trajectory is a spacelike curve ${\mathcal C}$ and is  illustrated in Fig.~\ref{fig:ABSTP} by the solid black line connecting $A$ to $B_\perp$. 

Fortunately , the optical-mechanical Eqns.~\ref{Helmholz:massless}-\ref{eq:force} already take into account general relativistic curvature effects within the geometrical optics approximation, and provides us with the Helmholtz equation in the spacelike hypersurface of $A$ and it  is written in the Cartesian coordinates of Alice's laboratory. The derivation of the paraxial wave equation, (Eq.~\ref{eq:sepe}) for optics from the Helmholtz equation (Eq.~\ref{eq:hez}) in  Sec.~\ref{sec:pwe} can be applied to this optical-mechanical Helmholtz equation (Eq.~\ref{Helmholz:massless}) with one modification. In flat spacetime the a plane-wave with wave number $k_0$ will propagate in a straight line in the laboratory with the same frequency. In a curved spacetime the photon will ordinarily be redshifted and its trajectory with tangent 3-vector, $k$, will ordinarily be curved in the laboratory 3-space.  For these reasons we would like to prove that we can transform the Helmoltz  given in Eq.~\ref{Helmholz:massless} at some point $P$ in in the $t=const$ spacelike hypersurface (e.g. Alice and Bob's laboratory)  as 
\begin{equation}
\label{eq:hezbar}
\left(\frac{\partial^2}{\partial \bar z^2} + \overline{\nabla}_T^2 \right) u + k^2 u =0,
\end{equation}
and $\overline{\nabla}_T$ is the transverse Laplacian in the barred coordinates,
\[
\overline{\nabla}_T^2 := \left(\frac{\partial^2}{\partial \bar x^2}\right) +\left(\frac{\partial^2}{\partial \bar y^2}\right).
\]
This ``barred'' frame is Fermi-Walker transported along the photon trajectory in the 3-space connecting $A$ to $B_\perp$ as illustrated in Fig.~\ref{fig:ABSTP}. 
Here $\bar z$ is defined at each point $P$ along ${\mathcal C}$ to be the proper spacelike distance along ${\mathcal C}$ from  $A$ to $P$, 
\begin{equation}
\bar z := \int_{\mathcal C} ds
\end{equation}
and the $\bar z$-axis, $e_{\bar z} = \partial_{\bar z}$, is the tangent to the curve ${\mathcal C}$ at each $P$ and is parallel to the spatial projection of the wave vector, 
\begin{equation}
k \cdot e_{\bar z} = 0,\ \forall P\in{\mathcal C}.
\end{equation}  
the transverse axes $e_{\bar x}$ and $e_{\bar y}$ are  Fermi-Walker transported along ${\mathcal C}$, 
\begin{equation}
\nabla_k e_{\bar i} = \nabla_{e_{\bar z}} e_{\bar i}  = 0,\  for\, \bar i=\bar x, \bar y, \bar z.
\end{equation} 
Whereas $\bar z$ measures the distance along $\mathcal C$ from $A$ to $P$, $\bar x$ and $\bar y$ measures the proper distance perpendicularly away from $\mathcal C$ along $e_{\bar x}$ and $e_{\bar y}$; respectively. The spatial part of $k$ is tangent to the photons trajectory in this photon trajectory frame. This as illustrated in in Fig.~\ref{fig:ABSTP} by the coordinate axes at event $P$.   A necessary condition for the derivation of the paraxial equation in this hypersurface is that we solve Eq.~\ref{eq:force} for the curve ${\mathcal C}$, and we have the Jacobian representing the rotational operator in space connecting the bared frame to the metric frame  
\begin{equation} 
\label{eq:omega}
e_i = \Omega^{\bar j}{}_{i} e_{\bar j} \ \ \longleftrightarrow  \ \  
\frac{\partial}{\partial x^i} = \frac{\partial \bar x^j}{ \partial x^i}  \frac{\partial}{\partial \bar x^j}
\end{equation}
and its inverse. 

Paralleling the development of the paraxial equation in  Sec.~\ref{sec:pwe}, we assume the field is the product of a slowly varying amplitude $\psi$  that serves to modulate a plane wave train traveling tangent to the wave vector. In our barred coordinate frame on ${\mathcal C}$, we let  
\begin{equation} 
\label{eq:psibar}
u = \psi\, e^{i k \bar z}.
\end{equation}
Here $u$, $\psi$ and $k$ all are functions of $\bar x$, $\bar y$ and $\bar z$. We begin to transform Eq.~\ref{Helmholz:massless} by Eq.~\ref{eq:omega},
\begin{align} 
\frac{\partial {\ }}{\partial {x^k}} & = \frac{\partial \bar x^i}{\partial x^k}  \frac{\partial}{\partial {\bar x^i}} = \Omega^{\bar i}{}_{k}  \frac{\partial}{\partial {\bar x^i}}\\
\frac{\partial^2 {\ }}{\partial {x^k}^2} & = \Omega^{\bar i}{}_{k}   \Omega^{\bar i}{}_{k} \frac{\partial^2}{\partial {\bar x^i}\partial {\bar x^j}}.
\end{align}
Inserting, Eq.~\ref{eq:psibar} into Eq.~\ref{Helmholz:massless} will yield the Helmholtz equation in the the barred coordinates. It is rather involved and we will not show it here. Furthermore, this equation will not easily be put into the form we assumed in Eq.~\ref{eq:hezbar}. In particular, we have mixed derivatives of $psi$  for an  arbitrary transformation matrix $\Omega$.  We believe we can justify approximations; however, this is beyond the scope of this manuscript and beyond the week-field application we intended.   Nevertheless, for weak gravitational fields, e.g. around the Earth, we can work in the isotropic coordinates without going to the bared coordinates. That is we expect minimal bending of the photon trajectories and the paraxial approximation about the $z$-axis should be valid to good approximation.  We then assume the transformation matrix is the identity matrix, 
\begin{equation}
\label{eq:OI}
\Omega = \mathbb{I}.
\end{equation} 
We then rewrite Eq.~\ref{Helmholz:massless} in the unbarred coordinates by singling out the $z$-axis for propagation, 
\begin{equation}
\left(\frac{\partial^2}{\partial z^2} + \nabla_T^2 \right) u + k^2 u =0,
\end{equation}
and $\nabla_T$ is the transverse Laplacian in the barred coordinates,
\[
\nabla_T^2 := \left(\frac{\partial^2}{\partial x^2}\right) +\left(\frac{\partial^2}{\partial y^2}\right).
\]
Again we parallel the development of the paraxial equation in  Sec.~\ref{sec:pwe} but this time for the coordinates $r=\{x,y,z\}$ of the $t=constant$ spacelike hypersurface of our conformally-flat spacetime.   We again assume that the  field is the product of a slowly varying amplitude $\psi$  that serves to modulate a plane wave train traveling tangent to the wave vector. In our global coordinates we again  let  
\begin{equation} 
u = \psi\, e^{i k z}.
\end{equation}
Here $u$, $\psi$ and $k$ all are functions of $x$, $y$ and $z$.  Since we are in isotropic flat coordinates, the $z$-direction is just as good as any direction we could have chosen.  Inserting Eq.~\ref{eq:psi} into Eq.~\ref{eq:hez}, we find
\begin{equation}
\begin{array}{ll}
&2 i  \left( 1 +  \frac{z}{k}\frac{\partial k}{\partial z }  \right) k  \frac{\partial \psi}{\partial z} 
+ \nabla_T^2 \psi 
-2 z k^2 \left(\frac{1}{k}\frac{\partial k}{\partial z } + \frac{z}{2}  \frac{\nabla k\cdot\nabla k}{k^2}\right) \psi  \\
&+ 2 i k \left(  \frac{1}{k}\frac{\partial k}{\partial z } + \frac{z}{2}  \frac{\nabla^2 k}{k}  \right)  \psi
+2 i z k \left( \frac{1}{k}\frac{\partial k}{\partial x}  \frac{\partial \psi}{\partial x}  +  \frac{1}{k}\frac{\partial k}{\partial y}  \frac{\partial \psi}{\partial y}  \right)
+ \frac{\partial^2 \psi}{\partial z ^2} = 0\,.
\end{array}
\end{equation}
We would like to prove that we can ignore the three terms in the second line of this equation, and we think that this is reasonable. However, we have not been able to provide a proof. We continue  with this assumption for the remainder of this manuscript. We will address the proof in a future paper. 
Therefore given these assumptions, we obtain the optical-mechanical equivalent of the paraxial equation for a beam propagating primarily down the $z$-axis in our spacetime coordinates,
\begin{equation}
\label{eq:thisisit} 
\boxed{
-\frac{1}{i}  \frac{\partial \psi}{\partial z} = -\frac{1}{2 \kappa} \nabla_T^2 \psi  + {\mathcal V} \psi\,,
}
\end{equation}
where the effective wave number is now 
\begin{equation}
\kappa =\left( 1 +  z\,\frac{1}{k}\frac{\partial k}{\partial z }  \right) k\,,
\end{equation}
and the effective potential is 
\begin{equation}
\label{eq:calV}
{\mathcal V} = -\frac{z k^2}{\kappa} 
 \left(\frac{1}{k}\frac{\partial k}{\partial z } + \frac{1}{2} z  \frac{\nabla k\cdot\nabla k}{k^2}\right)\,.
\end{equation}

This paraxial equation still predicts the bending of the photon trajectory and its redshift. Connecting with the previous section we generalized Eq.~\ref{eq:psi} and wrote, 
\begin{equation} 
\label{eq:psin}
u  = \psi_n\, e^{i S}.
\end{equation} 
where, the surface of constant phase  
\begin{equation}
S(r):=\int_{\mathcal C} k(r)\cdot dx = const,
\end{equation} 
has normals in the direction of  $\nabla S_0(r)=k(r)$ at each point $P$ along the photon's 3-trajectory connecting Alice to Bob. The wave 3-vector threads crossed each of  these phase surfaces at normal angles, and this defines the trajectory of our photon in the 3-space from Alice to Bob as illustrated in the left of Fig.~\ref{fig:ABSTP}. We will further this calculation for stronger gravitational bending effects by using the Fermi-Walker transported coordinates for Alice's lab in order to construct the paraxial equation from the Helmholtz equation.  This will be discussed in a future manuscript.

We are now in a position to simulate a photon's wavefront evolution in the optical-mechanical analogue using our split operator code to solve Eq.~\ref{eq:thisisit}. We provide a simple numerical example of the solution in the next section.

\section{SIMULATION OF AN OPTICAL VORTEX IN A CURVED SPACETIME}
\label{sec:misc}

We  simulate an optical vortex given by Eq.~\ref{eq:initialbeam} propagating outward along the $z$-axis in our isotropic coordinates of a Schwarzschild spacetime of mass $M$ given in conformally-flat coordinates. We start the simulation at $z=z_0$ and examine the the evolution using our split operator code. We numerically solve Eqs.~\ref{eq:thisisit}-\ref{eq:calV}.

\begin{figure} [ht]
\centering
   \includegraphics[height=2.0 in]{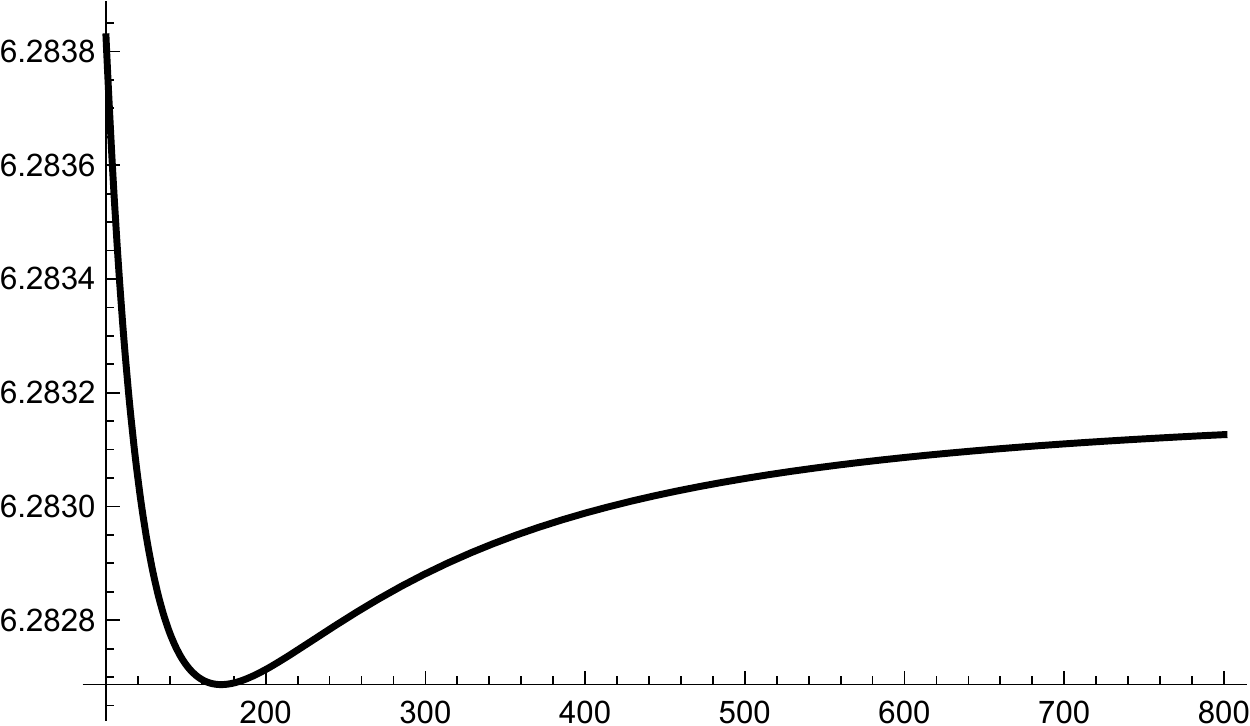}
   \caption
   { \label{fig:k}
This is the effective wavenumber, $\kappa$,  as a function of $z$ for $M=1$, $k_0=2 \pi$, $x=y=10$. We are most interested in the $z\gg 1$. Additionally, 
$ \lim_{x\rightarrow \infty} {\mathcal \kappa} = k_0=2\pi$, and $\lim_{x\rightarrow 0} {\mathcal \kappa}=8.29937$.
}
\end{figure}

\begin{figure} [ht]
   \centering
   \includegraphics[height=2.0 in]{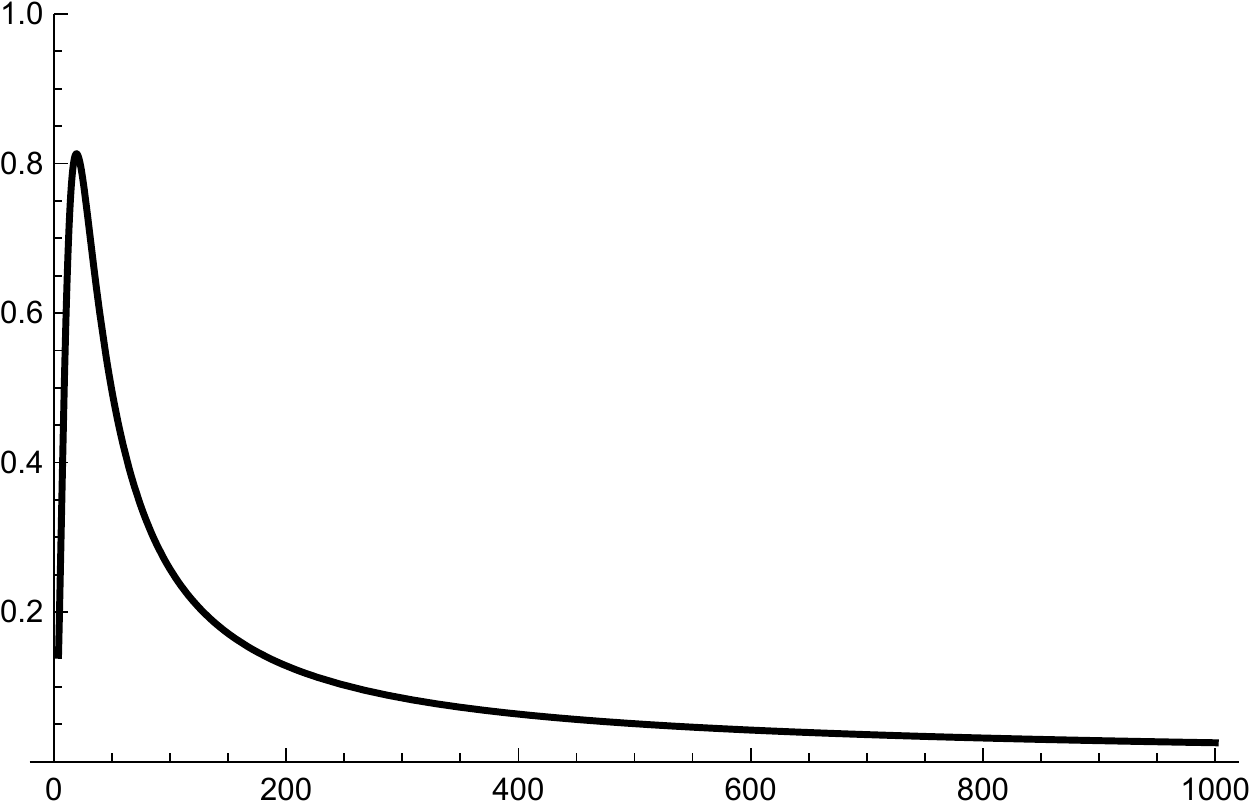}
     \caption[example]
   { \label{fig:V}
This is the effective potential, ${\mathcal V}$, as a function of $z$ for $M=1$, $k_0=2 \pi$, $x=y=10$. We are most interested in the $z\gg 1$. Additionally, 
$ \lim_{x\rightarrow \infty} {\mathcal V} = \lim_{x\rightarrow 0} {\mathcal V}=0$.
}
\end{figure}

In order to determine the effective wave number $\kappa$ and effective potential $\mathcal V$, we use the wave vector in Eq~\ref{Helmholz:massless} and Eq.~\ref{Sch:metric:iso}. We find, 
\begin{equation}
\kappa = \frac{k_0 \left(r+M\right)^2 \left(r^4 - 4 M r z^2 -M^2 \left(x^2+y^2-z^2\right)\right)}{r^4 \left(r-M\right)^2}
\end{equation}
and the effective potential
\begin{equation}
{\mathcal V} = 
\frac{2 k_0 M z^2 (M-2 r) (2 M-r) (M+r)^2}{r
   (M-r)^2 \left(M^2 \left(x^2+y^2-z^2\right)+4 M r
   z^2-r^4\right)} \, .
\end{equation}
  The effective wave number and the effective potential are  graphed in Figs.~\ref{fig:k}-\ref{fig:V}.

\begin{figure} [ht]
\centering
\includegraphics[height=2.5 in]{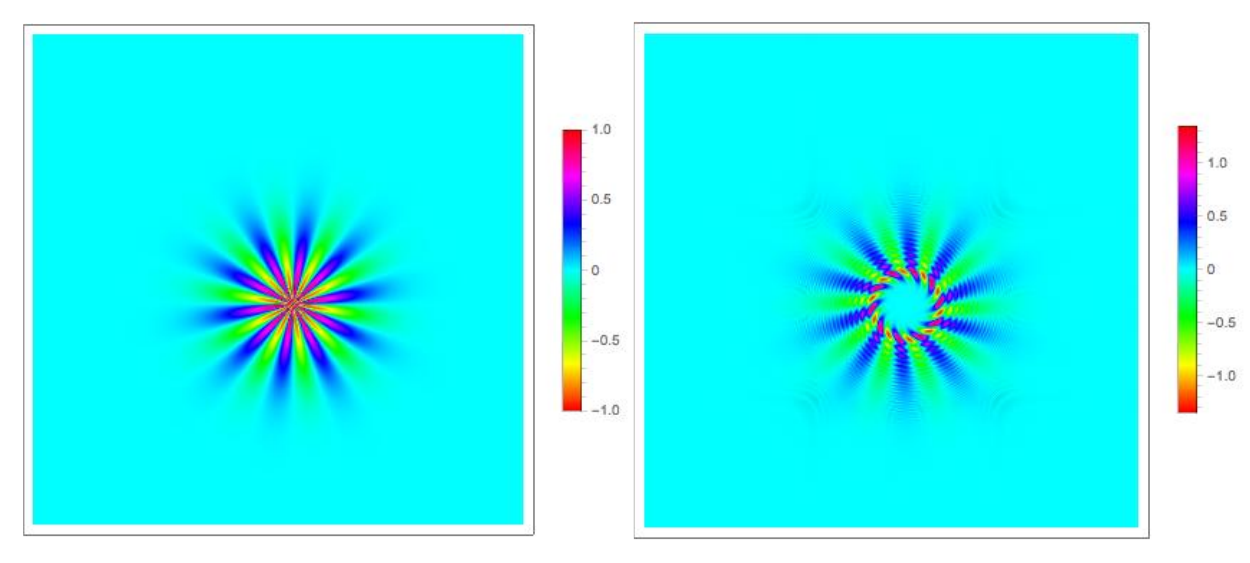}
\caption{
\label{fig:ovgr}
We display the evolution of a Gaussian optical vortex wavefront with $\sigma=4$  and $\ell=10$ in the conformally flat coordinates of a black hole of mass $M=1$ (in geometrical units).  The wavefront propagated parallel to the $z$-axis, but the peak of the initial Gaussian wavefront was offset in the $x$-axis at $x=2.5\lambda$ and $y=0$. We used $k_0=2\pi$ and $\lambda=1.0$. We simulated this evolution using $1024\times1024\times128$ in the $x$, 
$y$ and $z$ direction; respectively.  A plot of the real part of $\psi$ intensity of the initial optical vortex at $z_0=25$  is displayed to the left, and  the right plot displays the real part of the wavefront after it evolved inward toward $z=20$.  We see in the right figure the typical ringed hole along the propagating optical vortex and the phase wings emanating from this ring.    In addition to the effective wave number, $\kappa$ and effective potential ${\mathcal V}$'s effects on the wave front. We could also transform the wavefront into the usual Schwarzschild coordinates.
}
\end{figure}

We used the split operator code to propagate the $\ell=10$ optical vortex with a Gaussian envelope that we used in Sec.~\ref{sec:pwe}.  Here we set $M=1$,, $k_0=2\pi$ and started the wavefront  at $z_0=25$ and propagating inward toward the origin. We illustrate in Fig.~\ref{fig:ovgr} the initial wavefront intensity and the evolved wavefront showing the development of the orbital angular momentum  ($\ell=10$) ring. 

\section{CONCLUSION}
\label{sec:fini}

In this manuscript we reviewed the paraxial approximation and we outlined the split operator method to solve the paraxial equation.  We applied our numerical approach in flat Minkowski spacetime to numerically simulate a volume holographic sorter by propagating both a matched and mismatched optical vortex  through the glass. Using this example, we demonstrated the efficiency, accuracy and convergence of the code. We then numerically examined an optical-mechanical analogue model of the propagation of a photon in a curved spacetime.  We derived a curved-space paraxial wave equation  for this model that was derived from the optical-mechanical  Helmholtz equation.  While we have not yet formulated an paraxial evolution equation using the Fermi-Walker coordinates that track the trajectory of the photon, we are nevertheless able to construct a Schr\"odinger-like equation for a weak gravitational fields in the isotropic coordinates. More work is required in order to prove that two of the three terms we neglected are small compared to the other leading-order terms. We nevertheless, used the curved-spacetime paraxial equation to simulate an optical vortex propagating in a Schwarzschild spacetime that was expressed in an isotropic conformally-flat coordinate system.  We assumed that our two observers were co-moving synchronous observers to avoid further Lorentz boosts into an arbitrary rest frame, although this can be easily added by a Lorentz boost of our results.  

We look forward to reformulating this curved-spacetime paraxial approximation to simulate the scattering of wavefronts by a black hole at moderate impact parameters. This would require us to address the paraxial approximation in the Fermi-Walker coordinates where the wave vector is tangent to the world line of the photon in the  spacelike hypersurface.

\section*{ACKNOWLEDGMENTS}
PMA and WAM would like to acknowledge support of the Air Force Office of Scientific Research (AFOSR).  This work was supported under joint iC2S2 grants from the Korean Ministry of Science and Future Planning (MSIP) IITP 2017-0-00266 and the US Air Force Asian Office of Aerospace Research \& Development (AOARD)  FA2386-17-1-4070. Any opinions, findings, conclusions, or recommendations expressed in this material are those of the authors and do not necessarily reflect the views of AFRL.

\bibliography{Citations} 

\providecommand{\bysame}{\leavevmode\hbox to3em{\hrulefill}\thinspace}
\providecommand{\MR}{\relax\ifhmode\unskip\space\fi MR }
\providecommand{\MRhref}[2]{%
  \href{http://www.ams.org/mathscinet-getitem?mr=#1}{#2}
}
\providecommand{\href}[2]{#2}
\begin{thebibliography}{10}

\bibitem{Aksenov:08}
V.~P. Aksenov and Ch.~E. Pogutsa, \emph{Fluctuations of the orbital angular
  momentum of a laser beam, carrying an optical vortex, in the turbulent
  atmosphere}, Quantum Electronics \textbf{38} (2008), 343.

\bibitem{Allen:03}
L.~Allen, Stephen~M. Barnett, and Miles~J. Padgett, \emph{Optical angular
  momentum (optics \& optoelectronics)}, CRC Press, 2003.

\bibitem{Allen:92}
L.~Allen, M.~W. Beijersbergen, R.~J.~C. Spreeuw, and J.~P. Woerdman,
  \emph{Orbital angular momentum of light and the transformation of
  laguerre-gaussian laser modes}, Phys. Rev. A \textbf{45} (1992), 8185--8189.

\bibitem{Alsing:GRG:2001}
P.~M. Alsing, J.~C. Evans, and K.~K. Nandi, \emph{The phase of a quantum
  mechanical particle in curved spacetime}, Gen. Relativ. Gravit. \textbf{33}
  (2001), 1459--1487.

\bibitem{Bennett:84}
C.~H. Bennett and G.~Brassard, \emph{Quantum cryptography: public key
  distribution and coin tossing}, Proceedings of IEEE International Conference
  on Computers, Systems, and Signal Processing, Proc. IEEE, Bangalore, IEEE,
  New York, 1984, 1984, pp.~159--179.

\bibitem{Bruss:98}
Dagmar Bru\ss{}, \emph{Optimal eavesdropping in quantum cryptography with six
  states}, Phys. Rev. Lett. \textbf{81} (1998), 3018--3021.

\bibitem{Eddington:1920}
{Sir} Arthur~Stanley Eddington, \emph{Space, time and gravitation: an outline
  of the {General Relativity Theory}}, University Press, Cambridge, UK, 1920.

\bibitem{Ekert:00}
A.~Ekert, N.~Gisin, B.~Huttner, H.~Inamori, and H.~Weinfurter, \emph{Quantum
  cryptography}, The Physics of Information (D.~Bouwmeester, A.~Ekert, and
  A.~Zeilinger, eds.), Springer-Verlag Berlin Heidelberg, 2000, pp.~15--48;
  Ch.~2.

\bibitem{Ekert:91}
A.~K. Ekert, Phys. Rev. Lett. \textbf{67} (1991), 661.

\bibitem{Erhard2018}
Manuel Erhard, Robert Fickler, Mario Krenn, and Anton Zeilinger, \emph{Twisted
  photons: new quantum perspectives in high dimensions}, Light: Science \&Amp;
  Applications \textbf{7} (2018), 17146 EP --.

\bibitem{Cerf:02}
N.~J.~Cerf et. al., Phys. Rev. Lett. \textbf{88} (2002), 127902.

\bibitem{Alsing:AJP:2001}
J.~C. Evans, P.~M. Alsing, S.~Giorgetti, and K.~K. Nandi, \emph{Matter waves in
  a gravitational field: An index of refraction for massive particles in
  general relativity}, Am. J. Phys. \textbf{69} (2001), 1103--1110.

\bibitem{Evans:1996}
J.~C. Evans, K.~K. Nandi, and A.~Islam, \emph{On the optical-mechanical analogy
  in general relativity: Exact newtonian forms for the equations of motion of
  particles and photons}, Gen. Relativ. Gravit. \textbf{28} (1996), 413--439.

\bibitem{Evans:1986}
J.~C. Evans and M.~Rosenquist, \emph{F=ma optics}, Am. J. Phys. \textbf{54}
  (1986), 876--883.

\bibitem{Feit:1982}
M.~D. Feit, J.~A. Fleck, and A.~Steiger, \emph{Solution of the schr\"odinger
  equation by a spectral method}, J. Comput. Phys. \textbf{47} (1982), 412.

\bibitem{Feit:1980}
M.~D. Feit and J.~A.~Fleck Jr., Appl. Opt. \textbf{19} (1980), 1154,2240 and
  3140.

\bibitem{Hermann:1995}
M.~R. Hermann and J.~A. Fleck, \emph{Split-operator spectral method for solving
  the time-dependent schr\"odinger equation in spherical coordinates}, Phys.
  Rev. \textbf{A38} (1995).

\bibitem{Mair:01}
Alois Mair, Alipasha Vaziri, Gregor Weihs, and Anton Zeilinger,
  \emph{Entanglement of the orbital angular momentum states of photons}, Nature
  \textbf{412} (2001), 313 EP --.

\bibitem{Miller:84}
W.~A. Miller and J.~A. Wheeler, \emph{Delayed chouce experiments and bohr's
  elementary quantum phenomenon}, Foundations of Quantum Mechanics in the Light
  of New Technology (S.~Kamefuchi et~al., ed.), Physical Society of Japan,
  Tokyo, 1984, pp.~140--152.

\bibitem{Miller:2011}
Warner~A. Miller, \emph{Efficient photon sorter in a high-dimensional state
  space}, Quantum Info. Comput. \textbf{11} (2011), no.~3, 313--325.

\bibitem{Molina-Terriza:02}
Gabriel Molina-Terriza, Juan~P. Torres, and Lluis Torner, \emph{Management of
  the angular momentum of light: Preparation of photons in multidimensional
  vector states of angular momentum}, Phys. Rev. Lett. \textbf{88} (2001),
  013601.

\bibitem{Oemrawsingh:04}
S.~S.~R. Oemrawsingh, A.~Aiello, E.~R. Eliel, G.~Nienhuis, and J.~P. Woerdman,
  \emph{How to observe high-dimensional two-photon entanglement with only two
  detectors}, Phys. Rev. Lett. \textbf{92} (2004), 217901.

\bibitem{Paterson:05}
C.~Paterson, \emph{Atmospheric turbulence and orbital angular momentum of
  single photons for optical communication}, Phys. Rev. Lett. \textbf{94}
  (2005), 153901.

\bibitem{Groblacher:06}
et.~al. S.~Gl\"oblacher, New J. Phys. \textbf{8} (2006), 75.

\bibitem{Saleh:2007}
B.~A. Saleh and M.~C. Teich, \emph{Fundamentals of photonics}, 2nd ed., John
  Wiley \& Sons, Inc., Hoboken, NJ, 2007.

\end{thebibliography}
\bibliographystyle{amsplain} 

\end{document}